# Enhanced and robust superconductivity in $La_{0.8}Sr_{0.2}NiO_2$ membranes compressed up to 210 GPa

Shu Cai[1]*, Yuqing Tian[2]*, Shengjun Yan[2]*, Jinyu Zhao[1]*, Bo Hao[2], Jianfeng Zhang[1], Shuaihang Sun[1], Yang Ding[1], Qi Wu[3], Ho-kwang Mao[1,4], I. Božović[4], Yuefeng Nie[2,5]† and Liling Sun[1, 3]†

*[1]Center for High Pressure Science & Technology Advanced Research, Beijing 100193, China*

*[2]National Laboratory of Solid State Microstructures, Jiangsu Key Laboratory of Artificial Functional Materials, College of Engineering and Applied Sciences, Nanjing University, Nanjing 210093, China*

*[3]Institute of Physics, Chinese Academy of Sciences, Beijing 100190, China*

*[4]Shanghai Key Laboratory of Material Frontiers Research in Extreme Environments, Shanghai Advanced Research in Physical Sciences, Shanghai 201203, China*

*[5]Jiangsu Physical Science Research Center, Nanjing, 210093, China.*

The discovery of superconductivity in infinite-layer nickelate thin films has opened a new frontier for exploring unconventional oxide superconductors beyond the copper oxide family[1]. However, the presence of substrate hampers investigations under very high pressure. Recently, this problem was circumvented by a study of freestanding $Nd_{0.85}Sr_{0.15}NiO_2$ membranes, which revealed that superconducting transition temperature kept increasing as the pressure was ramped up to 91 GPa, without showing signs of saturation[2]. Naturally, one wonders what would happen at even higher pressures. Here, we report that superconductivity in a freestanding $La_{0.8}Sr_{0.2}NiO_2$ membrane persists under applied ultrahigh pressures up to 210 GPa. The superconducting transition onset temperature exhibits a continuous dome-like evolution, increasing from 16 K at ambient pressure to a peak of 74.5 K at 146 GPa, and then gradually decreasing to 57.4 K at 210 GPa. Such robustness of superconductivity against ultrahigh pressure has not been observed in high-$T_c$ oxide superconductors before[3-16].

After decades of search for superconductivity in nickelates, it was eventually discovered[1] in thin films of $Nd_{0.8}Sr_{0.2}NiO_2$, with the resistivity dropping fast below the "onset temperature" $T_c^{onset} \approx 17$ K. This has attracted much attention due to an analogy to cuprate superconductors, offering an opportunity to shed light on the origins of high-$T_c$ superconductivity[17-21]. Much like their cuprate counterparts, these infinite-layer nickelates adopt a layered crystal structure in which Ni assumes a $3d^9$ electronic configuration[22-27]. The normal state of optimally doped infinite-layer nickelate films also exhibits a linear dependence of resistivity on temperature, a hallmark of the strange metal observed in the optimally doped cuprate superconductors and several other unconventional superconducting materials[1,28-46]. These commonalities establish the infinite-layer nickelate films as valuable platform for investigating the microscopic origins of unconventional superconductivity. Subsequently, superconductivity was also reported in thin films of Ruddlesden-Popper (RP) nickelates such as $La_3Ni_2O_7$ (LNO), first only under high pressure[47], and subsequently also at the ambient pressure[48], and in Pr- or Sr- doped variants[49-54].

So far, infinite-layer and RP nickelate films were only grown on relatively thick substrates (typically 500 μm). This hampered measurements under high pressure in diamond anvil cells (DACs), where the height of the sample-hole is generally limited to less than 50 μm. However, a recent success in fabricating freestanding nickelate membranes offered a pathway to overcome this limitation, enabling measurements free from substrate-induced effects[55,56].

Applying pressure is a powerful and clean way of reducing interatomic distances, manipulating superconductivity[3,7,13,57] and unlocking new quantum phenomena[12,58,59]. Recent studies showed that the superconducting transition temperature in $Pr_{1-x}Sr_xNiO_2$ and $(La,Pr/Sr)_3Ni_2O_7$ films can be increased by applying pressure[60-62]. In particular, in a freestanding infinite-layer nickelate $Nd_{0.8}Sr_{0.2}NiO_2$, it increased up $T_c^{onset} \approx 74$ K at $P = 91.5$ GPa[2]. The enhancement of $T_c^{onset}$ followed a linear trend without any sign of saturation[2]. Several questions naturally arise: (1) Can superconductivity persist under ultrahigh pressures exceeding 100 GPa? (2) Will $T_c^{onset}$ rise further and up to what maximum? (3) What phenomena will emerge in freestanding membranes subjected to ultrahigh pressures? (4) Is the pressure response of $T_c$ in freestanding infinite-layer nickelates similar to those observed in cuprates and other high-$T_c$ oxide superconductors?

To address these issues, we applied high pressure to a freestanding $La_{0.8}Sr_{0.2}NiO_2$ membrane, whose behavior under such extreme conditions has not been explored before. The freestanding single-crystal $La_{0.8}Sr_{0.2}NiO_2$ thin films were synthesized following a procedure analogous to that reported previously[2,56]. Details about the sample synthesis are provided in the Methods section. Figure 1a schematically illustrates the epitaxial heterostructure configuration, highlighting the crystalline alignment of a $La_{0.8}Sr_{0.2}NiO_2$ layer deposited on a $SrTiO_3$-buffered substrate. To assess the sample quality, we first performed X-ray diffraction (XRD) measurements, scanned from $\theta$ to $2\theta$, on the $La_{0.8}Sr_{0.2}NiO_2$ thin films. As presented in Fig. 1b, the distinct (001) and (002) peaks corresponding to the infinite-layer phase with the $NiO_2$ planes parallel to the substrate surface, and absence of any discernible secondary phases. We subsequently measured the temperature-dependent resistance of $La_{0.8}Sr_{0.2}NiO_2$ both before and after the transfer process (Fig. 1c), and performed magnetic field sweeps with the field oriented perpendicular to the film. At zero field, a clear superconducting transition is observed, exhibiting a substantial resistance drop starting at $T_c^{onset} \approx 16$ K and reaching zero (i.e., the noise floor) resistance at around 12 K. Upon increasing magnetic fields up to 9 T, the transition temperature shifts to lower values monotonically, consistent with the systematic suppression of superconductivity.

Next, we conducted high-pressure resistance measurements on three membrane samples in a DAC. The transfer process of the $La_{0.8}Sr_{0.2}NiO_2$ membrane onto the diamond culet (the flat area of the diamond anvil) is illustrated in Fig. 2a. The details of the procedure are described in the Methods section. Following the transfer process, four electrodes were deposited onto the top surface of the membrane. Figure 2b presents a top-view optical image of the final device, in which the membrane is precisely covered by four gold electrodes, demonstrating excellent alignment between the membrane and the electrode pattern.

Our high-pressure resistance measurements were initially performed on sample 1 (S#1) using a DAC with 110 μm culet. As displayed in Fig. 2c, the sample exhibits metallic-like normal state behavior, followed by a superconducting transition with $T_c^{onset} \approx 18$ K at $P = 5$ GPa. Following the common practice in this field, and to enable comparison with Ref. 2 (and others) on equal footing, we first focus on this "onset" of the resistivity drop. To be specific, we define this $T_c^{onset}$ either by the 0.5% resistance drop ($T_c^{99.5\% R}$) or by the two-line intersection method $T_c^{2\text{-}line}$.

With increasing pressure, we observed a monotonic rise in $T_c^{onset}$. At $P = 100$ GPa, $T_c^{onset} \approx 66$ K. Remarkably, at $P = 146$ GPa, the maximum pressure attained in this experimental run, $T_c^{onset} \approx 74.5$ K. To the best of our knowledge, such extraordinary robustness of superconductivity against ultrahigh pressure has never been experimentally observed in any oxide superconducting system.

To verify the reproducibility of the observed high-pressure behavior and to explore whether $T_c$ can be further enhanced, we performed systematic electrical resistance measurements on sample 2 (S#2) over the pressure range of 26-165 GPa using a diamond anvil cell with a 100 μm culet. As shown in Fig. 2d, the $T_c^{onset}(P)$ evolution of S#2 basically follows the trend observed in S#1. In S#2 it reaches $T_c^{onset} \approx 72$ K at $P =$ 147 GPa and then exhibits a slight decrease. At $P = 165$ GPa, the sample still maintains $T_c^{onset} \approx 71$ K (Fig. 2d), confirming the robust nature of the superconducting state.

Consistent $T_c^{onset}(P)$ behavior was also observed in the third sample (S#3), which was measured over a pressure range of 35–210 GPa using a DAC with a 50 μm culet (Fig. 2e and 2f). We again observed that the transition onset shifts to higher temperatures with increasing pressure up to 147 GPa, above which it displays a slight decrease between 169 and 210 GPa (Fig. 2f). At $P = 174$ GPa, a pronounced resistance drop remains clearly visible. Upon further compression to $P = 210$ GPa, a resistance drop is still observed, with $T_c^{onset} \approx 57.4$ K.

Given that nickelate films exhibit broad superconducting transitions distinct from conventional BCS superconductors, a critical question arises: does the observed increase in $T_c^{onset}$ reflect a genuine enhancement of superconductivity, or is it merely a consequence of transition broadening that shifts the apparent onset temperature to higher values? To address this question, we take a derivative of the $R(T)$ dependence and determine $T_c^m$ as the temperature at which the d$R(T)$/d$T$ function reaches the maximum. We plot d$R(T)$/d$T$ curves for S#1 measured at different pressures and the $T_c^m(P)$ data in Fig. 3. The results show that $T_c^m$ increases with $P$. For example, $T_c^m = 13$ K at 15 GPa, while $T_c^m = 57$ K at $P = 147$ GPa, so $T_c^m$ increased by a factor of four. While $\delta T_c$ (the full width at half maximum of the peak in d$R(T)$/d$T$) indeed increases with $P$, this is a secondary effect – it is way too small to account for the observed increase in $T_c^{onset}$. Therefore, these results support our main conclusion that superconductivity in a freestanding $La_{0.8}Sr_{0.2}NiO_2$ membrane is enhanced under ultrahigh-pressure.

To further characterize superconductivity in freestanding $La_{0.8}Sr_{0.2}NiO_2$ membranes under extreme pressure, we applied different magnetic fields to the compressed sample 2 at $P = 80$ GPa. Figure 4a presents the temperature-dependent resistance under various magnetic fields. We found that the superconductivity is systematically suppressed upon increasing magnetic fields up to 7 T, along with a slight broadening of the offset. This is consistent with a superconducting transition.

In Figure 4b, we show a plot of $\mu_0 H_{c2}$ versus $T/T_c^m(H=0$ T$)$ obtained at the ambient pressure and at $P = 80$ GPa. The increase in $\mu_0 H_{c2}$ under pressure confirms the enhancement of superconductivity in compressed $La_{0.8}Sr_{0.2}NiO_2$. Our results are consistent with those obtained in $Nd_{0.85}Sr_{0.15}NiO_2$ membranes[2].

We have summarized our experimental results from three samples in the $P$-$T_c^{onset}$ phase diagram shown in Fig. 5. Superconductivity is found to persist over an exceptionally wide pressure range, from ambient pressure to the ultrahigh pressure of 210 GPa. The pressure effects on $T_c^{onset}$ for the three $La_{0.8}Sr_{0.2}NiO_2$ samples exhibit a dome-like behavior. $T_c^{onset}$ increases almost linearly below 60 GPa with a slope of 0.62

K/GPa. Above 60 GPa, it continues to increase with pressure but deviates downwards from the linear growth. It gradually reaches a maximum value at $P$ = 146 GPa, where for S#1, $T_c^{99.5\% R}$ = 74.5 K and $T_c^{2\ line}$ = 64 K, respectively.

Upon further compression, $T_c^{onset}$ exhibits a mild and gradual decrease. Even at $P$ = 210 GPa, the sample still maintains a remarkably high $T_c^{onset}$ = 57.4 K. At such high pressure, the overlap between Ni 3$d$ and O 2$p$ orbitals is expected to increase substantially, resulting in a great increase in electron kinetic energy. According to Mott physics picture, this would substantially weaken the effective Coulomb repulsion (reducing $U/t$). Nevertheless, we observe that at $P$ = 210 GPa, $T_c^{onset}$ still remains high. This phase diagram we inferred implies that the superconducting pairing strength in the nickelate 112 system remains robust and is not suppressed even across this wide pressure range.

In the phase diagram, we include $T_c^{onset}(P)$ data obtained from another freestanding infinite-layer superconductor, $Nd_{0.85}Sr_{0.15}NiO_2$[2] (open squares). We found that $T_c^{onset}$ values of both materials are enhanced by external pressure and increase linearly with a comparable slope below 60 GPa (Fig.5), implying that pressure exerts a similar effect on the electronic structure of these two materials. Above 60 GPa, however, their $T_c^{onset}(P)$ dependences start to diverge. For $Nd_{0.85}Sr_{0.15}NiO_2$, $T_c^{onset}$ continues to rise linearly and reaches a maximum value of $T_c^{onset}$ = 74.2 K at $P$ = 91.5 GPa[2]. In our $La_{0.8}Sr_{0.2}NiO_2$, the slope is a little lower than that of $Nd_{0.85}Sr_{0.15}NiO_2$ but $T_c^{onset}$ continues to rise nonlinearly up to $P$ = 146 GPa, where the highest $T_c^{onset}$ = 74.5 K is observed. (Here, we determined $T_c^{onset}$ is using the same method as described in Ref. 2). With further increasing pressure, the $T_c^{onset}$ of the $La_{0.8}Sr_{0.2}NiO_2$ membrane ceases to increase and shows a slow and steady decline (Fig.5). Additionally, we found that $T_c^{onset}$ observed in our samples is lower than that of $Nd_{0.85}Sr_{0.15}NiO_2$ under the same pressure conditions. This may be attributed to the differences in the lattice constants, given that the ionic radius of $La^{3+}$ is larger than that of $Nd^{3+}$, or to variations in doping level, which could optimize the state of $NiO_2$ plane. This warrants further investigation in future experiments. Nevertheless, although $La_{0.8}Sr_{0.2}NiO_2$ and $Nd_{0.85}Sr_{0.15}NiO_2$ attain their maximal transition temperatures at different pressures, both peak at $T_c^{onset}$ ≈ 75 K, implying that *f*-orbital magnetic moment has little impact on the enhancement of superconductivity, and that the $NiO_2$ planes instead govern superconductivity in infinite-layer nickelates.

In Fig. 5, the high-pressure behavior of the $La_{0.8}Sr_{0.2}NiO_2$ membrane is compared with that of the other high-$T_c$ oxide systems, including $(La,Pr/Sr)_3Ni_2O_7$ films[61,62]. Unlike the infinite-layer nickelate membrane, the $T_c^{onset}$ values of these reference systems decrease monotonically with increasing pressure above 7 GPa[62]. Additionally, superconductivity in optimally hole-doped $Bi_2Sr_2CaCu_2O_{8+\delta}$ (Bi-2212) and electron-doped $Pr_{0.87}LaCe_{0.13}CuO_{4+\delta}$ (PLCCO) is suppressed at approximately $P$ = 39 GPa (Ref. 12) and $P$ = 18 GPa (Ref. 63), respectively. A similar suppression trend is also observed in optimally doped $Ba_{0.6}K_{0.4}BiO_3$, where superconductivity disappears at $P$ = 20 GPa (Ref. 16). In contrast, superconductivity in the $La_{0.8}Sr_{0.2}NiO_2$ membrane persists from ambient pressure up to above 200 GPa (2 Mbar), an exceptionally robust behavior for a material that belongs to the typically fragile family of superconducting ceramics.

## Methods

### Sample Growth and Membrane Transfer

$La_{0.8}Sr_{0.2}NiO_3$ perovskite thin films were grown on (001)-oriented $TiO_2$-terminated $SrTiO_3$ substrates using a DCA R450 reactive molecular-beam epitaxy system. The water-soluble $Sr_4Al_2O_7$ sacrificial layer (3 unit cells) and a $SrTiO_3$ buffer layer (20 unit cells) were sequentially inserted between $La_{0.8}Sr_{0.2}NiO_3$ and $SrTiO_3$. An $SrTiO_3$ capping layer with a thickness of 6-to-13-unit cells was deposited on the surface of the $La_{0.8}Sr_{0.2}NiO_3$ thin films. $Sr_4Al_2O_7$ was grown at 850 °C under an oxygen partial pressure of 1×10−6 Torr, and $La_{0.8}Sr_{0.2}NiO_3$ was synthesized at 600 °C in an oxidant background pressure of 1×10−5 Torr using distilled ozone[56]. Subsequent topotactic reduction was conducted using 0.2 g $CaH_2$ as the reductant in a sealed quartz tube. The system was heated to 400 °C and kept at this temperature for 2 h, with the heating and cooling rates maintained at 10 °C·min$^{-1}$.

The freestanding $La_{0.8}Sr_{0.2}NiO_2$ membrane was chemically released by etching away the sacrificial layer in deionized water and fished out using a flexible polymer support matrix. Utilizing a high-precision micromanipulator transfer system, the micro-scale flexible membrane was precisely aligned and transferred onto the center of the diamond culet, followed by the patterning of micro-metallic electrodes to finalize the device assembly.

**High-Pressure Electrical Resistance Measurements**

*In situ* high-pressure electrical resistance measurements were systematically performed using diamond anvil cells. A metallic rhenium gasket was pre-indented and subsequently coated with an insulating layer of cubic boron nitride, in which a sample chamber hole was drilled. Gold electrodes were then deposited directly onto the the top of the sample to establish electrical contacts (Figure 2 and Supplementary Information). Diamond anvils with varying culet sizes were used to achieve different target pressure ranges. For measurements on samples S#1 and S#2, anvils with culet diameters of 110 and 100 μm were employed, respectively. To achieve ultrahigh-pressure conditions, sample S#3 was measured using a smaller culet diameter of 50 μm.

**Pressure Transmitting Medium and Determination**

Silicone oil was employed as the pressure-transmitting medium for all measurements. The sample pressure was determined using diamond Raman spectroscopy, with detailed information provided in the Supplementary Information.

*These authors contributed equally to this work.

Correspondence and requests for materials should be addressed to: Yuefeng Nie (ynie@nju.edu.cn) and Liling Sun (liling.sun@hpstar.ac.cn or llsun@iphy.ac.cn).

## Acknowledgements

The work was supported by the National Key Research and Development Program of China (Grants No. 2021YFA1401800, No. 2022YFA1403900, 2022YFA1402502 and 2021YFA1400400), National Natural Science Foundation of China (Grant No. 12434002) and Natural Science Foundation of Jiangsu Province (Grant No. BK20233001).

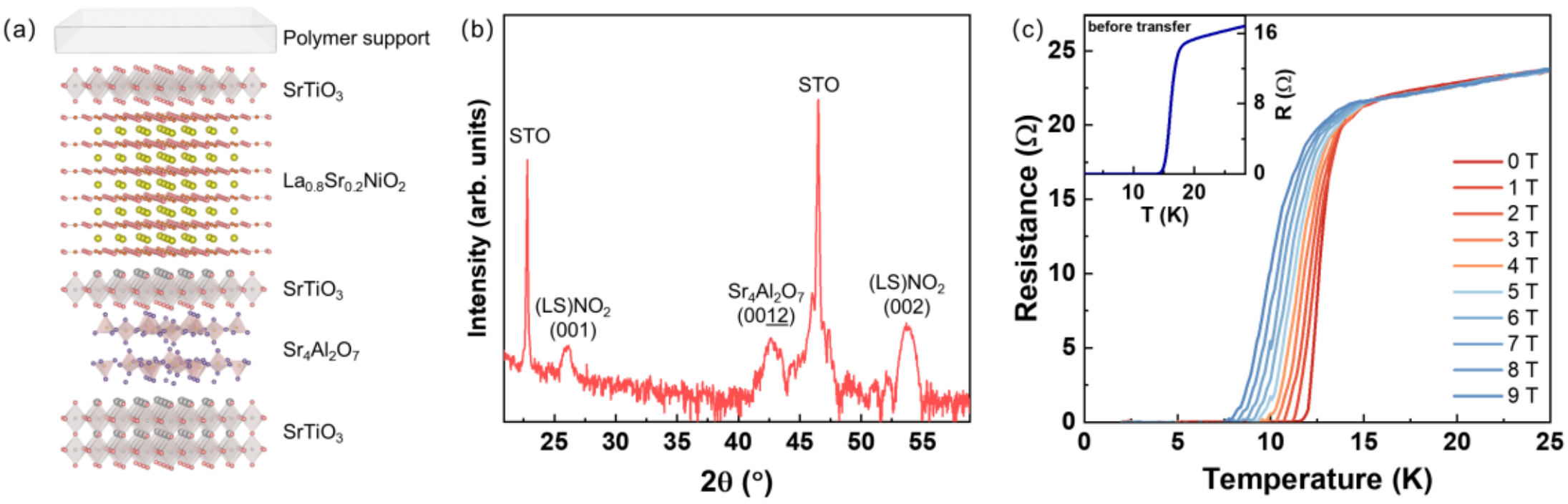

**Figure 1 Ambient-pressure characterization of the freestanding infinite-layer $La_{0.8}Sr_{0.2}NiO_2$ thin film.** (a) Schematic illustration of the epitaxial heterostructure stack prior to membrane exfoliation, depicting the integration of the infinite-layer nickelate film with the $SrTiO_3$ (STO) capping/buffer layers, as well as the water-soluble $Sr_4Al_2O_7$ sacrificial layer. (b) X-ray diffraction (XRD) $\theta$ - $2\theta$ scans of the infinite-layer phase assemblies on STO substrates. The presence of (001) and (002) reflections confirms a highly oriented out-of-plane crystalline texture. (c) Temperature-dependent resistance for the $La_{0.8}Sr_{0.2}NiO_2$ film, measured under external magnetic fields up to 9 T applied perpendicular to the *ab* plane. The inset shows the full-range resistance-versus-temperature curve acquired before the transfer process, confirming the intrinsic nature of the superconductivity.

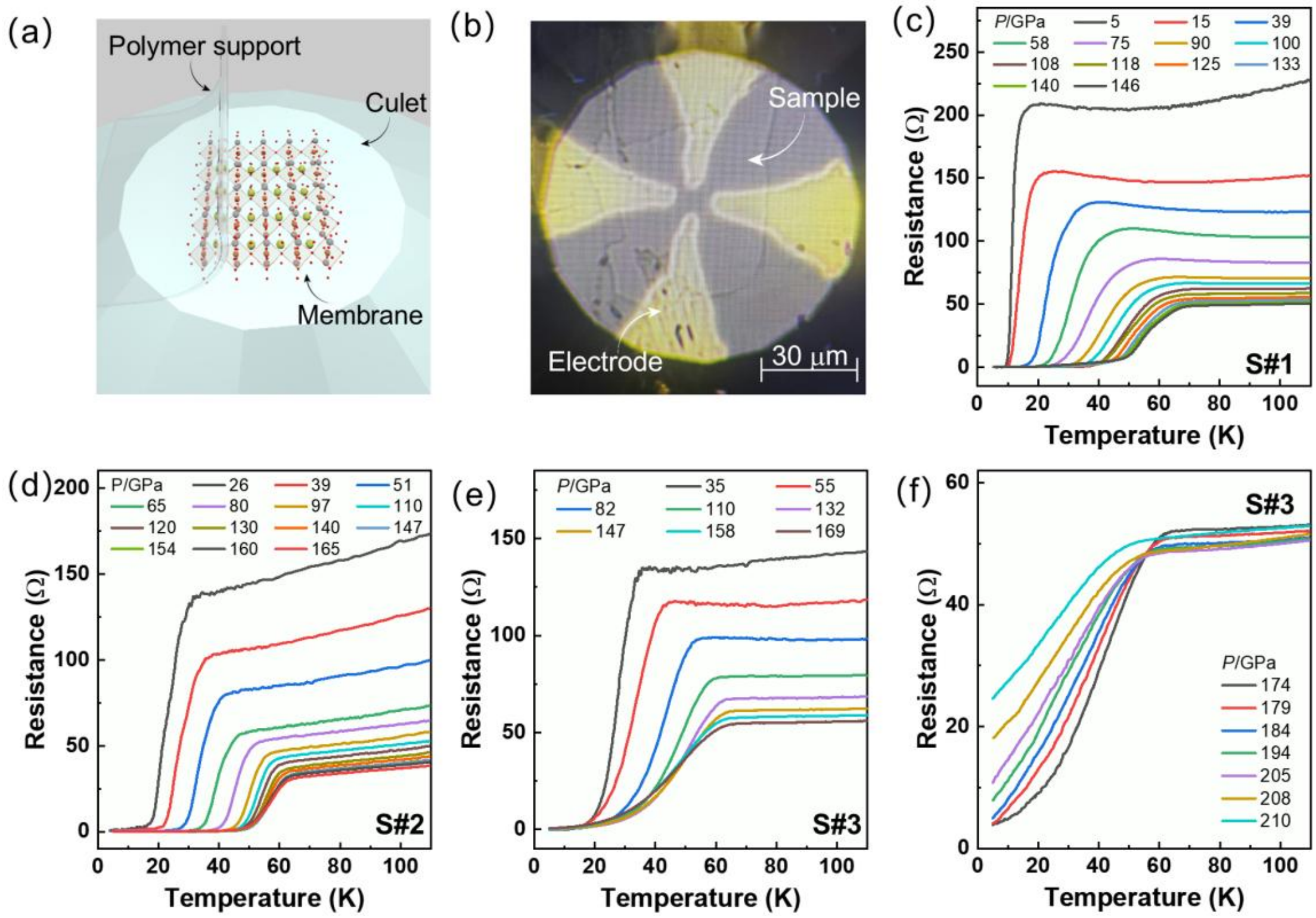

**Figure 2 The membrane and electrode configuration on the diamond culet, and temperature-dependent resistance plots at different pressures for the $La_{0.8}Sr_{0.2}NiO_2$ membranes.** (a) Schematic illustration of the membrane assembled on a diamond culet. (b) Top-view optical microscope image showing the $La_{0.8}Sr_{0.2}NiO_2$ membrane and the metallic electrodes. (c) Temperature-dependent resistance of sample S#1 measured for pressures ranging from 5 GPa to 146 GPa. (d) Resistance as a function of temperature for sample S#2 over the pressure range of 26-165 GPa. (e) Resistance versus temperature of sample S#3 across pressures from 35 to 169 GPa. (f) Data for S#3 measured in the ultrahigh pressure regime of 174-210 GPa.

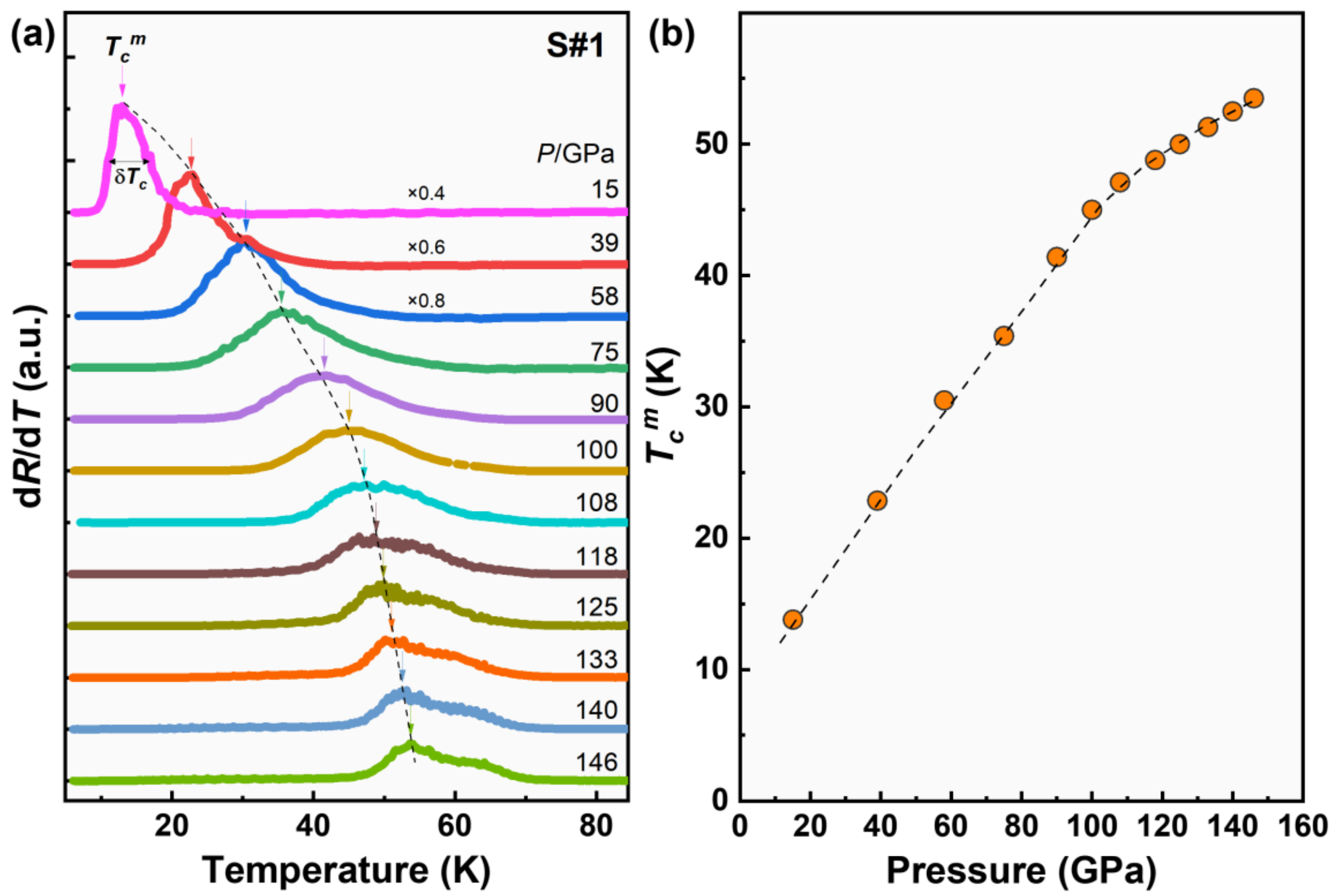

**Figure 3. Superconductivity enhancement in compressed $La_{0.8}Sr_{0.2}NiO_2$ membrane.** (a) Temperature derivative of resistance, d$R$($T$)/d$T$, for sample S#1 at various pressures. (b) Pressure dependence of the midpoint temperature of superconducting transition ($T_c^m$) for S#1, demonstrating an increase with increasing pressure.

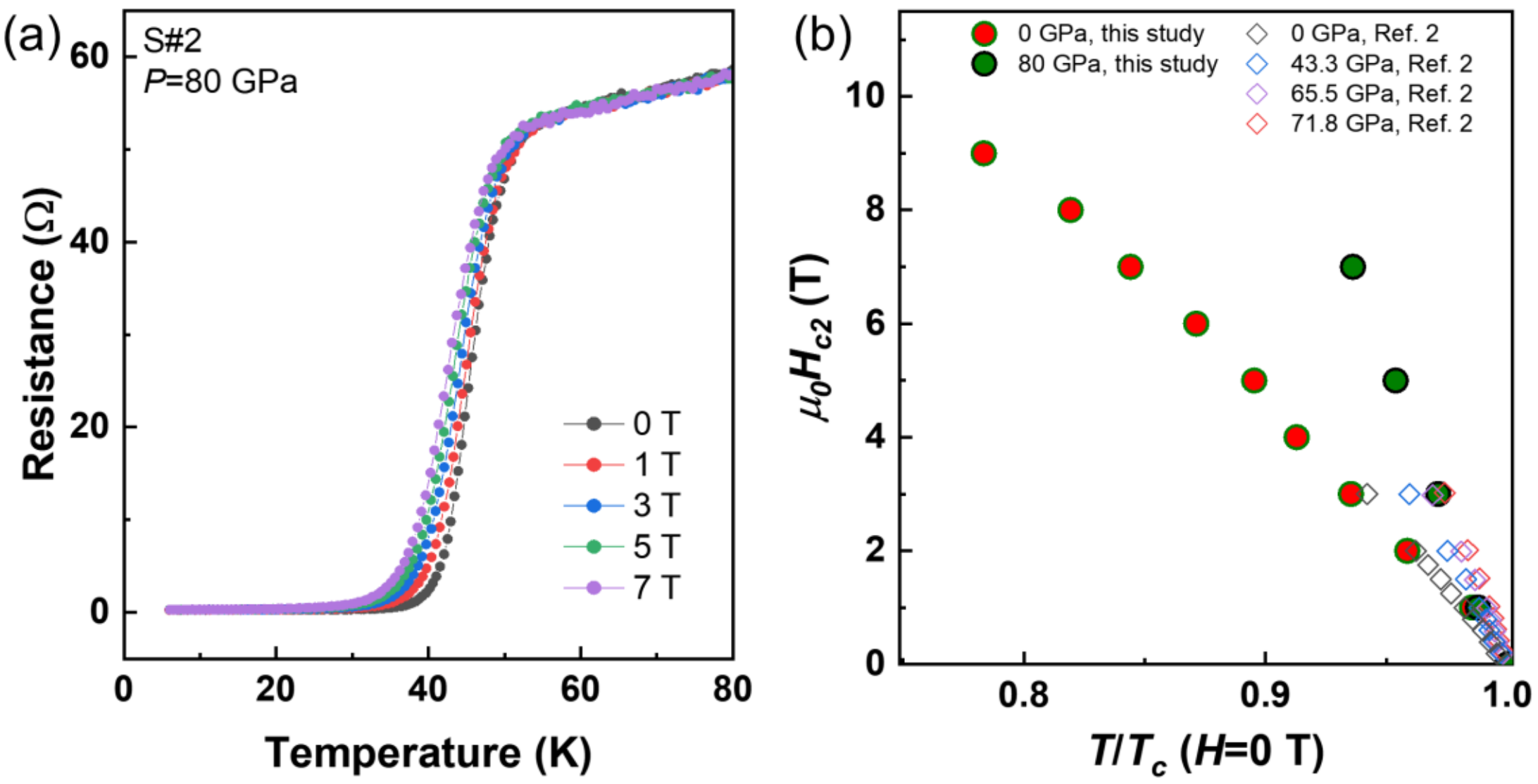

**Figure 4 Characterization of the superconducting transition and upper critical field of the $La_{0.8}Sr_{0.2}NiO_2$ membrane at high pressure.** (a) Temperature-dependent resistance under various magnetic fields for the membrane at 80 GPa. (b) The plot of $\mu_0 H_{c2}$ versus $T/T_c^m(H=0$ T$)$ obtained at the ambient pressure and at $P$ = 80 GPa (see solids). The open diamonds are the data taken from Ref. 2.

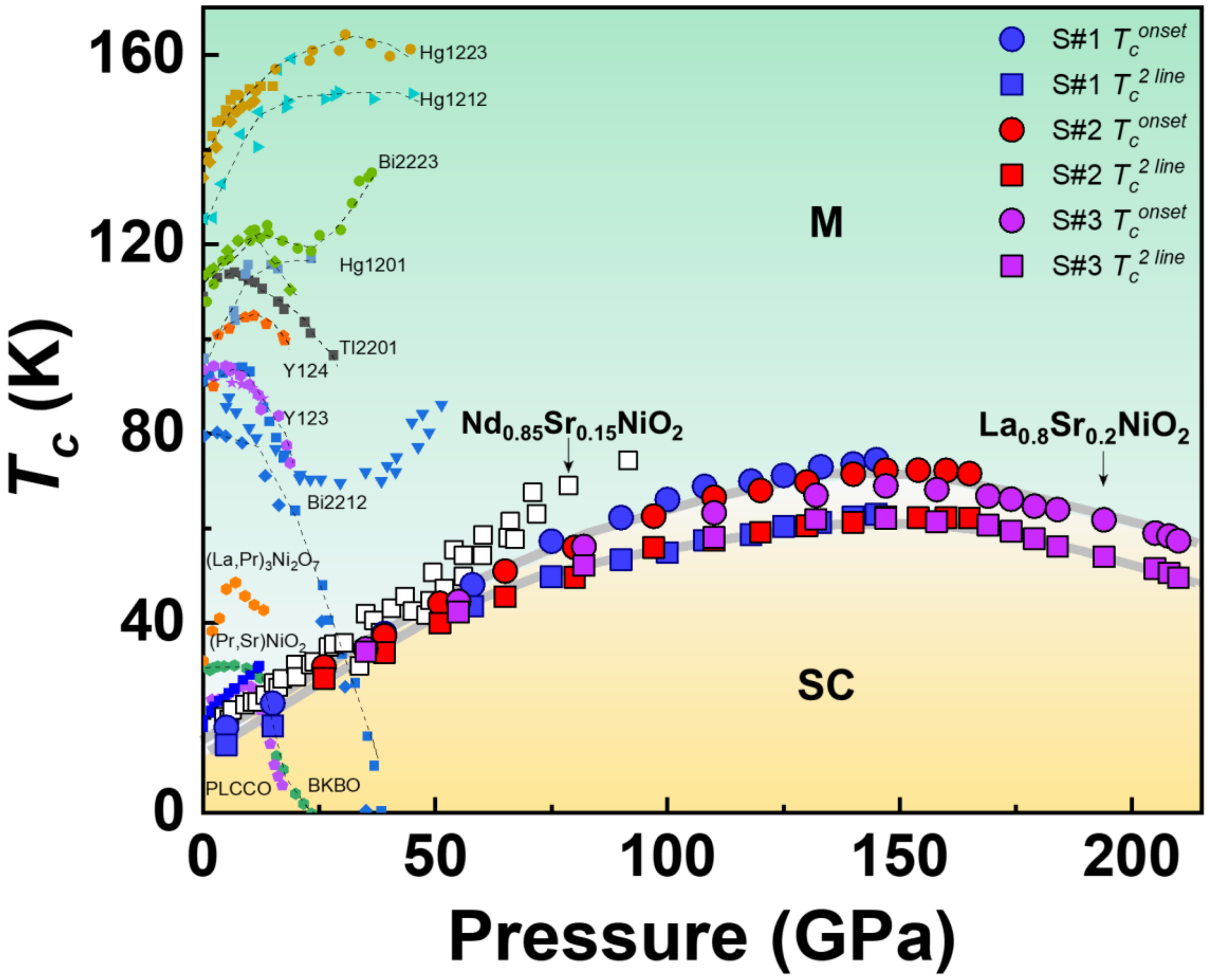

**Figure 5 Pressure-temperature phase diagram of $La_{0.8}Sr_{0.2}NiO_2$ membranes up to 210 GPa, along with comparative $T_c^{onset}(P)$ data for other superconducting systems.** The diagram summarizes the superconducting transition temperatures of three $La_{0.8}Sr_{0.2}NiO_2$ membranes as a function of applied pressure, alongside reported data for compressed $Nd_{0.85}Sr_{0.15}NiO_2$, (Pr,Sr)$NiO_2$, (La,Pr/Sr)$_3Ni_2O_7$, cuprate superconductors of $HgBa_2Ca_2Cu_3O_{8+\delta}$ (Hg1223), $HgBa_2CaCu_2O_{6+\delta}$ (Hg1212), $Bi_2Sr_2Ca_2Cu_3O_{10+\delta}$ (Bi2223), $HgBa_2CuO_{4+\delta}$ (Hg1201), $Tl_2Ba_2CuO_{6+\delta}$ (Tl2201), $YBa_2Cu_4O_8$ (Y124), $YBa_2Cu_3O_{7-\delta}$ (Y123), $Bi_2Sr_2CaCu_2O_{8+\delta}$ (Bi2212) and $B_{0.6}K_{0.4}BiO_3$ (BKBO). The solid circles (blue, red, and purple) represent data from this study, while white squares denote data taken from Ref. 2. The $T_c^{onset}$ (colored solid symbols) is determined by the 99.5% resistance drop, following the same criterion as in Ref. 2, whereas the $T_c^{2\ line}$ is determined by the two-line intersection method. The $T_c^{onset}(P)$ evolution of the $La_{0.8}Sr_{0.2}NiO_2$ membranes exhibits a dome-like behavior. In the low-pressure regime

below 60 GPa, $T_c^{onset}$ increases almost linearly with pressure with a slope of $dT_c^{onset}/dP \approx 0.62$ K/GPa. A maximum $T_c^{onset}$ of 74.5 K is reached at $P$ = 146 GPa, and the superconducting state persists up to $P$ = 210 GPa. For reference, comparative data are included for Hg1223 (Refs. 3-5), Hg1212 (Ref. 6), Bi2223 (Ref. 7), Hg1201 (Ref. 6), Tl2201 (Ref. 8), Y124(Ref. 9), Y123(Refs. 10,11), Bi2212 (Refs. 12-14), (Pr,Sr)$NiO_2$ (Ref. 60), (La,Pr/Sr)$_3Ni_2O_7$ (Refs. 61,62) and BKBO (Ref. 16).